\pdfoutput=1

\documentclass[sigconf,natbib,9pt]{acmart}

\copyrightyear{2017}
\setcopyright{acmlicensed}
\acmConference{ICTIR '17}{October 01-04, 2017}{Amsterdam, Netherlands}
\acmPrice{15.00}
\acmDOI{https://doi.org/10.1145/3121050.3121066}
\acmISBN{978-1-4503-4490-6/17/10}

\usepackage{anyfontsize}
\usepackage{amsmath}
\usepackage{amssymb}
\usepackage{booktabs}
\usepackage{color}
\usepackage{graphicx}
\usepackage{enumitem}
\usepackage{footnote}
\usepackage{grffile}
\usepackage{multirow}
\usepackage[all]{nowidow}
\usepackage[np, autolanguage]{numprint}
\usepackage{paralist}
\usepackage{setspace}
\usepackage{subfig}
\usepackage{tabularx}
\usepackage{todonotes}
\usepackage{flushend}

\setcitestyle{square,comma,numbers,sort&compress,sectionbib}

\hypersetup{pdfinfo={
Title={Structural Regularities in Text-based Entity Vector Spaces},
Author={Christophe Van Gysel, Maarten de Rijke, Evangelos Kanoulas}
}}

\newcommand{\BenchmarkTU}{TU}
\newcommand{\BenchmarkWThreeC}{W3C}

\newcommand{\Random}{random}
\newcommand{\GraphPCA}{Graph PCA}
\newcommand{\LSI}{LSI}
\newcommand{\LDA}{LDA}
\newcommand{\WordToVec}{word2vec}
\newcommand{\DocToVec}{doc2vec}
\newcommand{\SERT}{SERT}

\newcommand{\ResearchQuestionOne}{Do clusterings of text-based entity representations reflect the structure of their domains?}
\newcommand{\ResearchQuestionTwo}{To what extent do different text-based entity representation methods encode relations between entities?}

\newcommand{\NDCG}{NDCG}
\newcommand{\RPrec}{R-Precision}

\makeatletter
\newcolumntype{"}{@{\hskip\tabcolsep\vrule width 1pt\hskip\tabcolsep}}
\makeatother

\title{Structural Regularities in Text-based Entity Vector Spaces}

\author{Christophe Van Gysel}
\orcid{0000-0003-3433-7317}
\affiliation{%
\institution{University of Amsterdam}
\city{Amsterdam}
\country{The Netherlands}
}
\email{cvangysel@uva.nl}

\author{Maarten de Rijke}
\orcid{0000-0002-1086-0202}
\affiliation{%
\institution{University of Amsterdam}
\city{Amsterdam}
\country{The Netherlands}
}
\email{derijke@uva.nl}

\author{Evangelos Kanoulas}
\orcid{0000-0002-8312-0694}
\affiliation{%
\institution{University of Amsterdam}
\city{Amsterdam}
\country{The Netherlands}
}
\email{e.kanoulas@uva.nl}

\begin{document}

\begin{abstract}
Entity retrieval is the task of finding entities such as people or products in response to a query, based solely on the textual documents they are associated with. Recent semantic entity retrieval algorithms represent queries and experts in finite-dimensional vector spaces, where both are constructed from text sequences.

We investigate entity vector spaces and the degree to which they capture structural regularities. Such vector spaces are constructed in an unsupervised manner without explicit information about structural aspects. For concreteness, we address these questions for a specific type of entity: experts in the context of expert finding. We discover how clusterings of experts correspond to committees in organizations, the ability of expert representations to encode the co-author graph, and the degree to which they encode academic rank. We compare latent, continuous representations created using methods based on distributional semantics (LSI), topic models (LDA) and neural networks (word2vec, doc2vec, SERT). Vector spaces created using neural methods, such as doc2vec and SERT, systematically perform better at clustering than LSI, LDA and word2vec. When it comes to encoding entity relations, SERT performs best.
\end{abstract}

\ccsdesc[500]{Information systems~Content analysis and feature selection}

\maketitle


\section{Introduction}
\label{sec:introduction}

The construction of latent entity representations is a recurring problem \citep{Bordes2011,He2013learning,Zhao2015,Clark2016improving,Demartini2009} in natural language processing and information retrieval. So far, entity representations are mostly learned from relations between entities \citep{Bordes2011,Zhao2015} for a particular task in a supervised setting \citep{He2013learning}. How can we learn latent entity representations if
\begin{inparaenum}[(i)]
	\item entities only have relations to documents in contrast to other entities (e.g., scholars are represented by the papers they authored), and
	\item there is a lack of labeled data?
\end{inparaenum}

As entities are characterized by documents that consist of words, can we use word embeddings to construct a latent entity representation? Distributed representations of words \cite{Hinton1986}, i.e., word embeddings, are learned as part of a neural language model and have been shown to capture semantic \cite{Collobert2008} and syntactic regularities \cite{Mikolov2013word2vec,Pennington2014}. In addition, word embeddings have proven to be useful as feature vectors for natural language processing tasks \cite{Turian2010}, where they have been shown to outperform representations based on frequentist distributional semantics \cite{Baroni2014}. A downside of word embeddings \cite{Bengio2003} is that they do not take into account the document a word sequence occurred in or the entity that generated it.

\citet{Le2014} address this problem by extending \WordToVec{} models to \DocToVec{} by additionally modeling the document a phrase occurred in. That is, besides word embeddings they learn embeddings for documents as well. We can apply \DocToVec{} to the entity representation problem by representing an entity as a pseudo-document consisting of all documents the entity is associated with. Recent advances in entity retrieval incorporate real-world structural relations between represented entities even though the representations are learned from text only.  \citet{VanGysel2016experts} introduce a neural retrieval model (\SERT{}) for an entity retrieval task. In addition to word embeddings, they learn representations for entities.

In this paper, we study the regularities contained within entity representations that are estimated, in an unsupervised manner, from texts and associations alone. Do they correspond to structural real-world relations between the represented entities? E.g., if the entities we represent are people, do these regularities correspond to collaborative and hierarchical structures in their domain (industrial, governmental or academic organizations in the case of experts)? Answers to these questions are valuable because if they allow us to better understand the inner workings of entity retrieval models and give important insights into the entity-oriented tasks they are used for~\citep{Le2014}. In addition, future work can build upon these insights to extract structure within entity domains given only a document collection and entity-document relations so to complement or support structured information.

Our working hypothesis is that text-based entity representations encode regularities within their domain. To test this hypothesis
we compare latent text-based entity representations learned by neural networks (\WordToVec{}, \DocToVec{}, \SERT{}), count-based entity vector representations constructed using Latent Semantic Indexing (\LSI{}) and Latent Dirichlet Allocation (\LDA{}), dimensionality-reduced adjacency representations (\GraphPCA{}) and \Random{} representations sampled from a standard multivariate normal distribution. For evaluation purposes we focus on expert finding, a particular case of entity ranking. Expert finding is the task of finding the right person with the appropriate skills or knowledge~\cite{Balog2012}, based on a document collection and associations between people and documents. These associations can be extracted using entity linking methods or from document meta-data (e.g., authorship). Typical queries are descriptions of expertise areas, such as \emph{distributed computing}, and expert search engines answer the question ``Who are experts on \emph{distributed computing}?'' asked by people unfamiliar with the field.

Our main finding is that, indeed, semantic entity representations encode domain regularities. Entity representations can be used as feature vectors for clustering and that partitions correspond to structural groups within the entity domain. We also find that similarity between entity representations correlates with relations between entities. In particular, we show how representations of experts in the academic domain encode the co-author graph. Lastly, we show that one of the semantic representation learning methods, \SERT{}, additionally encodes importance amongst entities and, more specifically, the hierarchy of scholars in academic institutions. 

\section{Related work}

\subsection{Representations and regularities}

The idea that representations may capture linguistic and semantic regularities or even stereotyped biases that reflect everyday human culture has received considerable attention~\cite{Caliskan2017biases}. The idea of learning a representation of the elements of a discrete set of objects (e.g., words) is not new \cite{Rumelhart1985,Hinton1986}. However, it has only been since the turn of the last century that Neural Probabilistic Language Models (NPLM), which learn word embeddings as a side-effect of dealing with high-dimensionality, were shown to be more effective than Markovian models \cite{Bengio2003}. 

Even more recently, \citeauthor{Collobert2008} explain how the ideas behind NPLMs can be applied to arbitrary Natural Language Processing (NLP) tasks, by learning one set of word representations in a multi-task and semi-supervised setting. \citet{Turian2010} compare word representations learned by neural networks, distributional semantics and cluster-based methods as features in Named Entity Recognition (NER) and chunking. They find that both cluster-based methods and distributed word representations learned by NPLMs improve performance, although cluster-based methods yield better representations for infrequent words. \citet{Baroni2014} confirm the superiority of context-predicting (word embeddings) over context-counting (distributional semantics) representations.

Later algorithms are specifically designed for learning word embeddings \cite{Mikolov2013word2vec,Pennington2014}, such that, somewhat ironically, NPLMs became a side-product. These embeddings contain linguistic regularities~\cite{Mikolov2013regularities,Levy2014}, as evidenced in syntactic analogy and semantic similarity tasks. Multiple \emph{word} representations can be combined to form \emph{phrase} representations \cite{Mikolov2013compositionality}. Clusterings of word embeddings can be used to discover word classes \cite{Mikolov2013compositionality}. And insights gathered from word embedding algorithms can be used to improve distributional semantics~\citep{Levy2015}.

\subsection{Entity retrieval}

Around 40\% of web queries \cite{Pound2010objectretrieval} and over 90\% of academic search queries~\citep{Li2017academic} concern entities. Entity-oriented queries express an information need that is better answered by returning specific entities as opposed to documents \cite{Balog2010entitytrack}. The entity retrieval task is characterized by a combination of (noisy) textual data and semi-structured knowledge graphs that encode relations between entities \citep{Dietz2016tutorial}.

As a particular instance of entity retrieval, 
expert finding became popular with the TREC Enterprise Track \cite{TREC2010}. The task encompasses the retrieval of experts instead of documents. This is useful in enterprise settings, where employers seek to facilitate information exchange and stimulate collaboration~\citep{Davenport1998}. Expert finding diverges from the generic entity retrieval task due to the lack of explicit relations between experts. \citet{Balog2006} introduce language models for expert finding. In the maximum-likelihood language modeling paradigm, experts are represented as a normalized bag-of-words vector with additional smoothing. These vectors are high-dimensional and sparse due to the large vocabularies used in expert domains. Therefore, bag-of-words vectors are unsuited for use as representations as lower-dimensional and continuous vector spaces are preferred in machine learning algorithms \cite{Weber1998}. 
\citet{Demartini2009} introduce a framework for using document vector spaces in expert finding. \citet{Fang2010} explore the viability of learning-to-rank methods in expert retrieval.
\citet{vanDijk2015} propose methods for detecting potential experts in community question-answering.

\citeauthor{VanGysel2016experts} \cite{VanGysel2016products,VanGysel2016experts} propose a neural language modeling approach to expert finding; they also release the Semantic Entity Retrieval Toolkit (SERT) that we use in this paper. Closely related to expert finding is the task of expert profiling, of which the goal is to describe an expert by her areas of expertise \cite{Balog2007}, and similar expert finding \citep{Balog2007similarexperts}; see \cite{Balog2012} for an overview.

\subsection{Latent semantic information retrieval}

The mismatch between queries and documents is a critical challenge in search \citep{Li2014}. Latent Semantic Models (LSMs) retrieve objects based on conceptual, or semantic, rather than exact word matches. The introduction of Latent Semantic Indexing (LSI) \citep{Deerwester1990}, followed by probabilistic LSI (pLSI) \citep{Hofmann1999}, led to an increase in the popularity of LSMs. \citet{Salakhutdinov2009} perform unsupervised learning of latent semantic document bit patterns using a deep auto-encoder. \citeauthor{Huang2013} introduced Deep Structured Semantic Models \citep{Huang2013,Shen2014} that predict a document's relevance to a query using click data. Neural network models have also been used for learning to rank \citep{Burges2005,Deng2013,Liu2011}.


\section{Text-based Entity Vector Spaces}
\label{sec:methodology}

\newcommand{\Word}{w}
\newcommand{\Term}{t}

\newcommand{\Vocabulary}{V}

\newcommand{\Document}{d}
\newcommand{\Documents}{\MakeUppercase{\Document{}}}

\newcommand{\Query}{q}

\newcommand{\Candidates}{X}
\newcommand{\Candidate}{\MakeLowercase{\Candidates{}}}

\newcommand{\DocumentCandidates}[1][]{\Candidates{}_{\Document{}#1}}
\newcommand{\CandidateDocuments}[1][]{\Documents{}_{\Candidate{}#1}}

\newcommand{\Length}[1]{{|#1|}}

\newcommand{\WordRepresentation}{v}
\newcommand{\CandidateRepresentation}{e}

\newcommand{\CandidateBias}{b}

\newcommand{\VectorDim}{k}
\newcommand{\VectorMap}[1][]{g}

\newcommand{\ScoreFn}[2]{\text{score}(#1, #2)}

\newcommand{\Prob}[2][P]{#1(#2)}
\newcommand{\CondProb}[3][P]{\Prob[#1]{#2 \mid #3}}

\newcommand{\CosineSimilarity}[1]{S_C}

For text-based entity retrieval tasks we are given a document collection $\Documents{}$ and a set of entities $\Candidates{}$. Documents $\Document{} \in \Documents{}$ consist of a sequence of words $\Word{}_{1}, \ldots, \Word{}_\Length{\Document{}}$ originating from a vocabulary $\Vocabulary{}$, where $\Length{\cdot}$ denotes the document length in number of words. For every document $\Document{}$ we have a set $\DocumentCandidates{} \subset \Candidates{}$ of associated entities ($\DocumentCandidates{}$ can be empty for some documents) and conversely $\CandidateDocuments{} \subset \Documents{}$ consists of all documents associated with entity $\Candidate{}$. The associations between documents and experts can be obtained in multiple ways. E.g., named-entity recognition can be applied to the documents and mentions can subsequently be linked to entities. Or associations can be extracted from document meta-data (e.g., authorship).

Once determined, the associations between entities $\Candidates{}$ and documents $\Documents{}$ encode a bipartite graph. If two entities $\Candidate{}_i, \Candidate{}_j \in \Candidates{}$ are associated with the same document, we say that $\Candidate{}_i$ and $\Candidate{}_j$ are co-associated. However, the semantics of a co-association are equivocal as the semantics of an association are ambiguous by itself (e.g., author vs. editor). Therefore, instead of relying solely on document associations, we use the textual data of associated documents to construct an entity representation.

Vector space models for document retrieval, such as LSI \cite{Deerwester1990} or LDA \cite{Blei2003}, can be adapted to entity retrieval. We substantiate this for a specific entity retrieval task: expert finding. As there are many more documents than experts, it is not ideal to estimate a vector space directly on the expert-level using bag-of-word vectors (e.g., by representing every expert as a concatenation of its documents) due to data sparsity. Therefore, it is preferable to first estimate a vector space on the document collection and then use the obtained document representations to construct an entity vector. \citet{Demartini2009} take an entity's representation to be the sum of its documents:
\begin{equation}
\label{eq:expert-vsm}
\CandidateRepresentation{}_i = \sum_{\Document{}_j \in \CandidateDocuments[_i]{}} \VectorMap{}({\Document{}_j}),
\end{equation}
where $\CandidateRepresentation{}_i$ is the $\VectorDim{}$-dimensional vector representation of entity $\Candidate{}_i \in \Candidates{}$ and $\VectorMap{}$ is the function mapping a document to its vector space representation (e.g., \LSI{}). The dimensionality $\VectorDim{}$ depends on the underlying vector space. For simple bag-of-words representations, $\VectorDim{}$ is equal to the number of words in the vocabulary. For latent vector spaces (e.g., LSI), the $\VectorDim{}$-dimensional space encodes latent concepts and the choice of $\VectorDim{}$ is left to the user.

Vector space models for document retrieval are often constructed heuristically. E.g., Eq.~\ref{eq:expert-vsm} does not make optimal use of document-entity associations as document representations are added without taking into consideration the significance of words contained within them~\cite{Luhn1958}. And if many diverse documents are associated with an expert, then Eq.~\ref{eq:expert-vsm} is likely to succumb to the noise in these vectors and yield meaningless representations.

To address this problem, \citet{Le2014} introduced \DocToVec{} by adapting the \WordToVec{} models to incorporate the document a phrase occurs in. They optimize word and document embeddings jointly to predict a word given its context and the document the word occurs in. The key difference between \WordToVec{} and \DocToVec{} is that the latter considers an additional meta-token in the context that represents the document. Instead of performing dimensionality reduction on bag-of-words representations, \DocToVec{} learns representations from word phrases. Therefore, we use the \DocToVec{} model to learn expert embeddings by representing every expert $\Candidate{}_j \in \Candidates{}$ as a pseudo-document consisting of the concatenation of their associated documents $\CandidateDocuments[_j]{}$.

A different neural language model architecture than \DocToVec{} was proposed by \citet{VanGysel2016experts}, specifically for the expert finding task. For a given word $\Word{}_i$ and expert $\Candidate{}_j$:
\begin{equation}
\label{eq:expert-loglinear}
\ScoreFn{\Word{}_i}{\Candidate{}_j} = \exp{\left(\WordRepresentation{}_i^\intercal \cdot \CandidateRepresentation{}_j + \CandidateBias{}_j\right)},
\end{equation}
\newcommand{\ProbCandidateGivenWord}{\CondProb{\Candidates{} = \Candidate{}_j}{\Word{}_i}}%
where $\WordRepresentation{}_i$ ($\CandidateRepresentation{}_j$, resp.) are the latent $\VectorDim$-dimensional representations of word $\Word{}_i$ (and expert $\Candidate{}_j$, respectively) and $\CandidateBias{}_j$ is the bias scalar associated with expert $\Candidate{}_j$. Eq.~\ref{eq:expert-loglinear} can be interpreted as the unnormalized factor product of likelihood $\CondProb{\Word{}_i}{\Candidate{}_j}$ and prior $\Prob{\Candidate{}_j}$ in log-space. The score is then transformed to the conditional probability
\[
\ProbCandidateGivenWord{} = \frac{\ScoreFn{\Word{}_i}{\Candidate{}_j}}{\sum_{\Candidate{}_l \in \Candidates{}} \ScoreFn{\Word{}_i}{\Candidate{}_l}}.
\]
Unlike Eq.~\ref{eq:expert-vsm}, the conditional probability distribution $\ProbCandidateGivenWord{}$ will be skewed towards relevant experts if the word $\Word{}_i$ is significant as described by \citet{Luhn1958}. The parameters $\WordRepresentation{}_i$, $\CandidateRepresentation{}_j$ and $\CandidateBias{}_j$ are learned from the corpus using gradient descent. See \cite{VanGysel2016experts} for details.

Our focus lies on representations of entities $\CandidateRepresentation{}_j$ and how these correspond to structures within their domains (i.e., organizations for experts). These representations are estimated using a corpus only and can be interpreted as vectors in word embedding space that correspond to entities (i.e., people) instead of words. 


\section{Experimental set-up}

\subsection{Research questions}

\newcommand{\RQ}[2]{
	\begin{description}
	\item[\fontsize{10pt}{\baselineskip}\selectfont RQ#1] #2
	\end{description}
}

We investigate regularities within text-based entity vector spaces, using expert finding as our concrete test case, and ask how these representations correspond to structure in their respective domains. We seek to answer the following research questions:
\RQ{1}{\ResearchQuestionOne{}}
Many organizations consist of smaller groups, committees or teams of experts who are appointed with a specific role. When we cluster expert representations, do the clusters correspond to these groups?
\RQ{2}{\ResearchQuestionTwo{}}
The associations within expert domains encode a co-association graph structure. To what extent do the different expertise models encode this co-association between experts? In particular, if we rank experts according to their nearest neighbors, how does this ranking correspond to the academic co-author graph?

\subsection{Expert finding collections}
\label{sec:collections}

We use publicly-available expert finding collections provided by the World Wide Web Consortium (\BenchmarkWThreeC{}) and Tilburg University (\BenchmarkTU{}); see Table~\ref{tbl:collections}.

\newcommand{\specialcell}[3][c]{\begin{tabular}[#1]{@{}#2@{}}#3\end{tabular}}

\begin{table*}[t]
\centering
	\caption{An overview of the two expert finding collections (\BenchmarkWThreeC{} and \BenchmarkTU{}).\label{tbl:collections}}
	\smallskip

	\resizebox{0.60\textwidth}{!}{%
	\def\arraystretch{1.1125}
	\begin{tabular}{lc@{$\,$}lc@{$\,$}lc@{$\,$}l}
	\toprule
	\multicolumn{1}{c}{} & \multicolumn{2}{c}{\BenchmarkWThreeC{}} & \multicolumn{2}{c}{\BenchmarkTU{}} \\
	\midrule
	Documents in collection & \numprint{331037}\phantom{.00} & & \phantom{0}\numprint{31209}\phantom{.00} &  \\
	Average tokens per document & \phantom{00}\numprint{1237.23} & & \phantom{00}\numprint{2454.93} & \\

	\\

	Number of candidate experts & \phantom{000,}\numprint{715}\phantom{.00} & & \phantom{000,}\numprint{977}\phantom{.00} & \\

	\\

	\specialcell{l}{Number of document-candidate associations} & \numprint{200939}\phantom{.00} & & \phantom{0}\numprint{36566}\phantom{.00} & \\
	\specialcell{l}{Number of documents (with $\Length{\DocumentCandidates{}} > 0$)} & \phantom{0}\numprint{93826}\phantom{.00} & & \phantom{0}\numprint{27834}\phantom{.00} &  \\
	\specialcell{l}{Number of associations per document} & \phantom{000,00}\numprint{2.14} & $\pm\, 3.29$ & \phantom{000,00}\numprint{1.13} & $\pm\, 0.39$ \\
	\specialcell{l}{Number of associations per candidate} & \phantom{000,}\numprint{281.03} & $\pm\, 666.63$ & \phantom{000,0}\numprint{37.43} & $\pm\, 61.00$ \\

	\bottomrule
	\end{tabular}}
\end{table*}%

\smallskip

\smallskip\noindent%
\textbf{\BenchmarkWThreeC{}}. The \BenchmarkWThreeC{} collection was released as part of the 2005--2006 editions of the TREC Enterprise Track \cite{Craswell2005}. It contains a heterogeneous crawl of W3C's website (June 2004) and consists of mailing lists and discussion boards among others. In the 2005 edition, TREC released a list of working groups and their  members. Each working group is appointed to study and report on a particular aspect of the World Wide Web to enable the W3C to pursue its mission. We use the associations provided by \citet{VanGysel2016experts}, which they gathered by applying named entity recognition and linking these mentions to a list of organization members, as proposed by \citet{Balog2006}.

\smallskip\noindent
\textbf{\BenchmarkTU{}}. The \BenchmarkTU{} collection consists of a crawl of a university's internal website and contains bi-lingual documents, such as academic publications, course descriptions and personal websites \citep{Berendsen2013}. The document-candidate associations are part of the collection. For every member of the academic staff, their academic title is included as part of the collection.

\subsection{Implementations and parameters}

We follow a similar experimental set-up as previous work \cite{Balog2006,Demartini2009,Mikolov2013compositionality,VanGysel2016experts}. For \LSI{}, \LDA{}, \WordToVec{} and \DocToVec{} we use the Gensim\footnote{\url{https://radimrehurek.com/gensim}} implementation, while for the log-linear model we use the Semantic Entity Retrieval Toolkit\footnote{\url{https://github.com/cvangysel/SERT}} (\SERT{}) \citep{VanGysel2017SERT}. 

The corpora are normalized by lowercasing and removing punctuation and numbers. The vocabulary is pruned by removing stop words and  retaining the 60k most frequent words. We sweep exponentially over the vector space dimensionality ($\VectorDim{} = 32,\, 64,\, 128$ and $256$) of the methods under comparison. This allows us to evaluate the effect of differently-sized vector spaces and their modeling capabilities. 

For \WordToVec{}, a query/document is represented by its average word vector, which is effective for computing short text similarity \cite{Kenter2015}. We report both Continuous Bag-of-Words (CBOW) and Skip-gram (SG) variants of \WordToVec{}. 
For \LDA{}, we set $\alpha = \beta = 0.1$ and train the model for 100 iterations or until topic convergence is achieved. Unlike \citet{VanGysel2016experts}, for \SERT{}, we do not initialize with pre-trained \WordToVec{} embeddings. Default parameters are used in all other cases.

For \LSI{}, \LDA{} and \WordToVec{}, expert representations are created from document representations according to Eq.~\ref{eq:expert-vsm}. 

In addition to text-based representations, we also include two baselines that do not consider textual data. For the first method (\GraphPCA{}), we construct a weighted, undirected co-association graph where the weight between two entities is given by the number of times they are co-associated. We then apply Principal Component Analysis to create a latent representation for every entity. Secondly, we include a baseline where experts are represented as a \Random{} vector sampled from a standard multivariate normal distribution.


\newcommand{\LatentModels}{\Random{}, \GraphPCA{}, \LSI{}, \LDA{}, \WordToVec{}, \DocToVec{} and \SERT{}}

\newcommand{\Significant}{^{*\phantom{**}}}
\newcommand{\MoreSignificant}{^{**\phantom{*}}}
\newcommand{\HighlySignificant}{^{***}}

\section{Regularities in Entity Vector Spaces}

\newcommand{\RQRef}[1]{{RQ#1}}

We investigate regularities within latent text-based entity vector spaces. In particular, we first build latent representations for experts and ground these in the structure of the organizations where these experts are active. First, we cluster latent expert representations using different clustering techniques and compare the resulting clusters to committees in a standards organization of the World Wide Web (\RQRef{1}). We continue by investigating to what extent these representations encode entity relations (\RQRef{2}). We complement the answers to our research questions with an analysis of the prior (the scalar bias in Eq.~\ref{eq:expert-loglinear}) associated with every expert in one of the models we consider, \SERT{}, and compare this to their academic rank.

\subsection{Answers to research questions}
\newcommand{\RQAnswer}[3]{
    \begin{description}
    \item[\fontsize{10pt}{0}\selectfont RQ#1] {\fontsize{10pt}{0}\selectfont \textbf{#2}}%
    \end{description}%
    { #3 }
}

\smallskip
\RQAnswer{1}{\ResearchQuestionOne{}}{%
\newcommand{\NumClusters}{K}
\begin{figure*}[!t]
    \centering
    \includegraphics[width=0.85\textwidth]{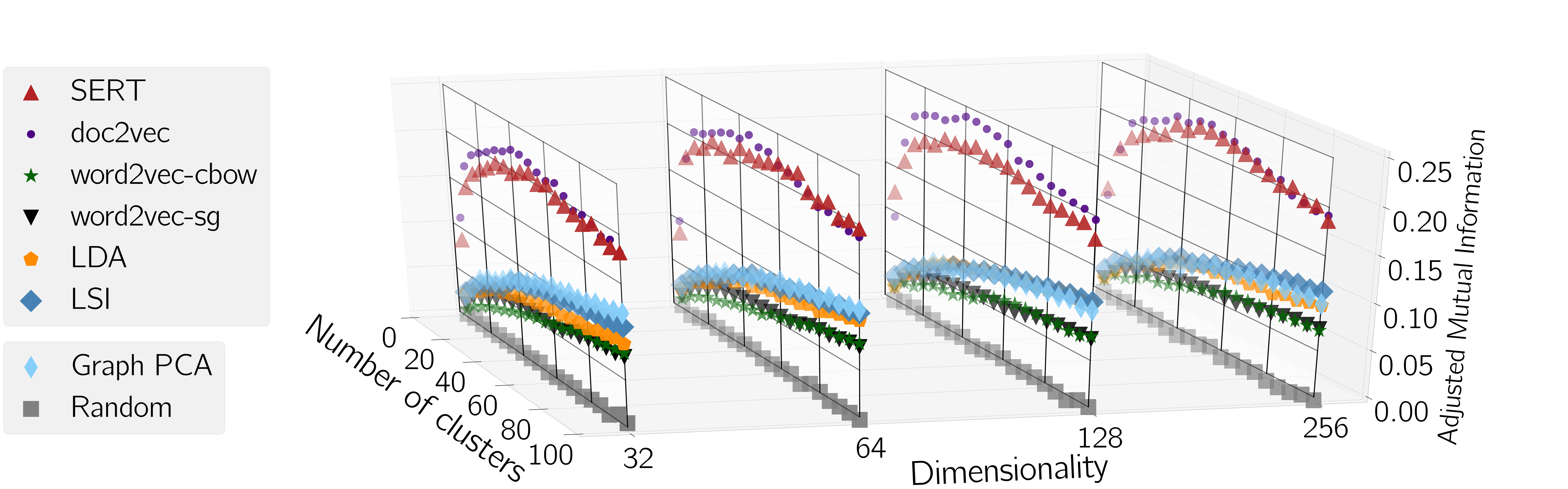}
    \smallskip
    \caption{Comparison of clustering capabilities of expert representations (\LatentModels{}) using $\NumClusters{}$-means for $10^0 \leq \NumClusters{} < 10^2$ (y-axis). The x-axis shows the dimensionality of the representations and the z-axis denotes the Adjusted Mutual Information.\label{fig:clustering}}
    \bigskip
\end{figure*}
\noindent%
The World Wide Web Consortium (\BenchmarkWThreeC{}) consists of various working groups.\footnote{\url{http://www.w3.org/Consortium/activities}}
Each working group is responsible for a particular aspect of the WWW and consists of two or more experts. We use these working groups as ground truth for evaluating the ability of expert representations to encode similarity. The \BenchmarkWThreeC{} working groups are special committees that are established to produce a particular deliverable \citep[p. 492]{Robert2011rules} and are a way to gather experts from around the organization who share areas of expertise and who would otherwise not directly communicate. Working groups are non-hierarchical in nature and represent clusters of experts. Therefore, they can be used to evaluate to what extent entity representations can be used as feature vectors for clustering.

We cluster expert representations using $\NumClusters{}$-means \cite{MacQueen1967}. While $\NumClusters{}$-means imposes strong assumptions on cluster shapes (convexity and isotropism), it is still very popular today due to its linear time complexity, geometric interpretation and absence of hard to choose hyper-parameters (unlike spectral variants or DBSCAN). We cluster expert representations of increasing dimensionality $\VectorDim{}$ ($\VectorDim{} = 2^i$ for $5 \leq i < 9$) using a linear sweep over the number of clusters $\NumClusters{}$ ($10^0 \leq \NumClusters{} < 10^2$).

During evaluation we transform working group memberships to a hard clustering of experts by assigning every expert to the smallest working group to which they belong as we wish to find specialized clusters contrary to general clusters that contain many experts. We then use Adjusted Mutual Information, an adjusted-for-chance variant of Normalized Information Distance~\cite{Vinh2010}, to compare both clusterings. Adjusting for chance is important as non-adjusted measures (such as BCubed precision/recall\footnote{This can be verified empirically by computing BCubed measures for an increasing number of random partitions.} as presented by \citet{Amigo2009}) have the tendency to take on a higher value for a larger value of $\NumClusters{}$. Performing the adjustment allows us to compare clusterings for different values of $\NumClusters{}$. We repeat the $\NumClusters{}$-means clustering 10 times with different centroids initializations and report the average.

Figure~\ref{fig:clustering} shows the clustering capabilities of the different representations for different values of $\NumClusters{}$ and vector space dimensionality. Ignoring the random baseline, representations built using \WordToVec{} perform worst. This is most likely due to the fact that document representations for \WordToVec{} are constructed by averaging individual word vectors. Next up, we observe a tie between \LSI{} and \LDA{}. Interestingly enough, the baseline that only considers entity-document associations and does not take into account textual content, \GraphPCA{}, outperforms all representations constructed from document-level vector space models (Eq.~\ref{eq:expert-vsm}). Furthermore, \DocToVec{} and \SERT{} perform best, regardless of vector space dimensionality, and consistently outperform the other representations. If we look at the vector space dimensionality, we see that the best clustering is created using 128-dimensional vector spaces. Considering the number of clusters, we see that \DocToVec{} and \SERT{} peak at about 40 to 60 clusters. This corresponds closely to the number of ground-truth clusters. The remaining representations (\WordToVec{}, \LSI{}, \LDA{}, \GraphPCA{}) only seem to plateau in terms of clustering performance at $\NumClusters{} = 100$, far below the clustering performance of the \DocToVec{} and \SERT{} representation methods. 

To answer our first research question, we conclude that expert representations can be used to discover structure within organizations. However, the quality of the clustering varies greatly and use of more advanced methods (i.e., \DocToVec{} or \SERT{}) is recommended.
}

\RQAnswer{2}{\ResearchQuestionTwo{}}{
\begin{table*}[!t]
    \caption{Retrieval performance (\NDCG{} and \RPrec{}) when ranking experts for a query expert by the cosine similarity of expert representations (\LatentModels{}) for the \BenchmarkTU{} expert collection (\S\ref{sec:collections}) for an increasing representation dimensionality. The relevance labels are given by the number of times two experts were co-authors of academic papers. Significance of results is determined using a two-tailed paired Student t-test ($^{*} \, p < 0.10$, $^{**} \, p < 0.05$, $^{***} \, p < 0.01$) between the best performing model and second best performing method.
    \label{tbl:graph}}
    \centering
    \bigskip
    \resizebox{0.80\textwidth}{!}{%
    \begin{tabular}{l c c c c c c c c}%
\toprule%
$\text{Dimensionality } \VectorDim  = $&\multicolumn{2}{c}{32}&\multicolumn{2}{c}{64}&\multicolumn{2}{c}{128}&\multicolumn{2}{c}{256}\\%
\cmidrule(lr){2-3}
\cmidrule(lr){4-5}
\cmidrule(lr){6-7}
\cmidrule(lr){8-9}
&NDCG&R{-}Precision&NDCG&R{-}Precision&NDCG&R{-}Precision&NDCG&R{-}Precision\\%
\midrule%
Random&\nprounddigits{2}\npdecimalsign{.}\numprint{0.177236415094}$\phantom{\HighlySignificant}$&\nprounddigits{2}\npdecimalsign{.}\numprint{0.00996264150943}$\phantom{\HighlySignificant}$&\nprounddigits{2}\npdecimalsign{.}\numprint{0.177525660377}$\phantom{\HighlySignificant}$&\nprounddigits{2}\npdecimalsign{.}\numprint{0.00726377358491}$\phantom{\HighlySignificant}$&\nprounddigits{2}\npdecimalsign{.}\numprint{0.177086037736}$\phantom{\HighlySignificant}$&\nprounddigits{2}\npdecimalsign{.}\numprint{0.00762226415094}$\phantom{\HighlySignificant}$&\nprounddigits{2}\npdecimalsign{.}\numprint{0.179706603774}$\phantom{\HighlySignificant}$&\nprounddigits{2}\npdecimalsign{.}\numprint{0.0103903773585}$\phantom{\HighlySignificant}$\\%
Graph PCA&\nprounddigits{2}\npdecimalsign{.}\numprint{0.380876226415}$\phantom{\HighlySignificant}$&\nprounddigits{2}\npdecimalsign{.}\numprint{0.179528679245}$\phantom{\HighlySignificant}$&\nprounddigits{2}\npdecimalsign{.}\numprint{0.387263018868}$\phantom{\HighlySignificant}$&\nprounddigits{2}\npdecimalsign{.}\numprint{0.196906792453}$\phantom{\HighlySignificant}$&\nprounddigits{2}\npdecimalsign{.}\numprint{0.411703962264}$\phantom{\HighlySignificant}$&\nprounddigits{2}\npdecimalsign{.}\numprint{0.233318113208}$\phantom{\HighlySignificant}$&\nprounddigits{2}\npdecimalsign{.}\numprint{0.386152641509}$\phantom{\HighlySignificant}$&\nprounddigits{2}\npdecimalsign{.}\numprint{0.232906603774}$\phantom{\HighlySignificant}$\\%
\midrule
LSI&\nprounddigits{2}\npdecimalsign{.}\numprint{0.386723584906}$\phantom{\HighlySignificant}$&\nprounddigits{2}\npdecimalsign{.}\numprint{0.167349245283}$\phantom{\HighlySignificant}$&\nprounddigits{2}\npdecimalsign{.}\numprint{0.431202641509}$\phantom{\HighlySignificant}$&\nprounddigits{2}\npdecimalsign{.}\numprint{0.20766490566}$\phantom{\HighlySignificant}$&\nprounddigits{2}\npdecimalsign{.}\numprint{0.460599811321}$\phantom{\HighlySignificant}$&\nprounddigits{2}\npdecimalsign{.}\numprint{0.225984150943}$\phantom{\HighlySignificant}$&\nprounddigits{2}\npdecimalsign{.}\numprint{0.473885849057}$\phantom{\HighlySignificant}$&\nprounddigits{2}\npdecimalsign{.}\numprint{0.227925471698}$\phantom{\HighlySignificant}$\\%
LDA&\nprounddigits{2}\npdecimalsign{.}\numprint{0.437495849057}$\phantom{\HighlySignificant}$&\nprounddigits{2}\npdecimalsign{.}\numprint{0.186046603774}$\phantom{\HighlySignificant}$&\nprounddigits{2}\npdecimalsign{.}\numprint{0.446252075472}$\phantom{\HighlySignificant}$&\nprounddigits{2}\npdecimalsign{.}\numprint{0.199429245283}$\phantom{\HighlySignificant}$&\nprounddigits{2}\npdecimalsign{.}\numprint{0.464408490566}$\phantom{\HighlySignificant}$&\nprounddigits{2}\npdecimalsign{.}\numprint{0.219798867925}$\phantom{\HighlySignificant}$&\nprounddigits{2}\npdecimalsign{.}\numprint{0.523460754717}$\phantom{\HighlySignificant}$&\nprounddigits{2}\npdecimalsign{.}\numprint{0.277198301887}$\phantom{\HighlySignificant}$\\%
word2vec{-}sg&\nprounddigits{2}\npdecimalsign{.}\numprint{0.463946037736}$\phantom{\HighlySignificant}$&\nprounddigits{2}\npdecimalsign{.}\numprint{0.222067735849}$\phantom{\HighlySignificant}$&\nprounddigits{2}\npdecimalsign{.}\numprint{0.486528679245}$\phantom{\HighlySignificant}$&\nprounddigits{2}\npdecimalsign{.}\numprint{0.238368490566}$\phantom{\HighlySignificant}$&\nprounddigits{2}\npdecimalsign{.}\numprint{0.493312264151}$\phantom{\HighlySignificant}$&\nprounddigits{2}\npdecimalsign{.}\numprint{0.237186981132}$\phantom{\HighlySignificant}$&\nprounddigits{2}\npdecimalsign{.}\numprint{0.498984528302}$\phantom{\HighlySignificant}$&\nprounddigits{2}\npdecimalsign{.}\numprint{0.250447924528}$\phantom{\HighlySignificant}$\\%
word2vec{-}cbow&\nprounddigits{2}\npdecimalsign{.}\numprint{0.459394716981}$\phantom{\HighlySignificant}$&\nprounddigits{2}\npdecimalsign{.}\numprint{0.225746603774}$\phantom{\HighlySignificant}$&\nprounddigits{2}\npdecimalsign{.}\numprint{0.472065849057}$\phantom{\HighlySignificant}$&\nprounddigits{2}\npdecimalsign{.}\numprint{0.238029056604}$\phantom{\HighlySignificant}$&\nprounddigits{2}\npdecimalsign{.}\numprint{0.478136037736}$\phantom{\HighlySignificant}$&\nprounddigits{2}\npdecimalsign{.}\numprint{0.24816490566}$\phantom{\HighlySignificant}$&\nprounddigits{2}\npdecimalsign{.}\numprint{0.482222830189}$\phantom{\HighlySignificant}$&\nprounddigits{2}\npdecimalsign{.}\numprint{0.253563018868}$\phantom{\HighlySignificant}$\\%
doc2vec&\nprounddigits{2}\npdecimalsign{.}\numprint{0.346980754717}$\phantom{\HighlySignificant}$&\nprounddigits{2}\npdecimalsign{.}\numprint{0.138771320755}$\phantom{\HighlySignificant}$&\nprounddigits{2}\npdecimalsign{.}\numprint{0.357944716981}$\phantom{\HighlySignificant}$&\nprounddigits{2}\npdecimalsign{.}\numprint{0.15120490566}$\phantom{\HighlySignificant}$&\nprounddigits{2}\npdecimalsign{.}\numprint{0.362356415094}$\phantom{\HighlySignificant}$&\nprounddigits{2}\npdecimalsign{.}\numprint{0.160383773585}$\phantom{\HighlySignificant}$&\nprounddigits{2}\npdecimalsign{.}\numprint{0.352983584906}$\phantom{\HighlySignificant}$&\nprounddigits{2}\npdecimalsign{.}\numprint{0.145858679245}$\phantom{\HighlySignificant}$\\%
SERT&\nprounddigits{2}\npdecimalsign{.}\textbf{\numprint{0.526321698113}}$\HighlySignificant$&\nprounddigits{2}\npdecimalsign{.}\textbf{\numprint{0.288272830189}}$\HighlySignificant$&\nprounddigits{2}\npdecimalsign{.}\textbf{\numprint{0.544672075472}}$\HighlySignificant$&\nprounddigits{2}\npdecimalsign{.}\textbf{\numprint{0.311938679245}}$\HighlySignificant$&\nprounddigits{2}\npdecimalsign{.}\textbf{\numprint{0.534917735849}}$\HighlySignificant$&\nprounddigits{2}\npdecimalsign{.}\textbf{\numprint{0.303620566038}}$\HighlySignificant$&\nprounddigits{2}\npdecimalsign{.}\textbf{\numprint{0.532707169811}}$\phantom{\HighlySignificant}$&\nprounddigits{2}\npdecimalsign{.}\textbf{\numprint{0.306999245283}}$\Significant$\\%
\bottomrule%
\end{tabular}}
\end{table*}

\noindent%
The text-based entity representation problem is characterized by a bipartite graph of entities and documents where an edge denotes an entity-document association. This differs from entity finding settings where explicit entity-entity relations are available and fits into the scenario where representations have to be constructed from unstructured text only. If latent text-based entity representations encode co-associations, then we can use this insight for
\begin{inparaenum}[(1)]
    \item a better understanding of text-based entity representation models, and
    \item the usability of latent text-based entity representations as feature vectors in scenarios where relations between entities are important.
\end{inparaenum}

We evaluate the capacity of text-based expert representations to encode co-associations by casting the problem as a ranking task. Contrary to typical expert finding, where we rank experts according to their relevance to a textual query, for the purpose of answering \RQRef{2}, we rank experts according to their cosine similarity w.r.t.\ a query expert \citep{Balog2007similarexperts}. This task shares similarity with content-based recommendation based on unstructured data \citep{Pazzani2007recsys}.

In expert finding collections, document-expert associations can indicate many things. For example, in the \BenchmarkWThreeC{} collection, entity-document associations are mined from expert mentions \citep{Balog2006}. However, for the \BenchmarkTU{} collection, we know that a subset of associations corresponds to academic paper authorship. Therefore, we construct ranking ground-truth from paper co-authorship and take the relevance label of an expert to be the number of times the expert was a co-author with the query expert (excluding the query expert themselves). Our intention is to determine to what extent latent entity representations estimated from text can reconstruct the original co-author graph. Given that we estimate the latent entity representations using the complete \BenchmarkTU{} document collection, by design, our evaluation is contained within our training set for the purpose of this analysis.

Table~\ref{tbl:graph} shows \NDCG{} and \RPrec{} \cite[p.~158]{Manning2008ir} for various representation models and dimensionality. \SERT{} performs significantly better than the other representations methods (except for the $256$-dimensional representations where significance was not achieved w.r.t.\ \LDA{}). \SERT{} is closely followed by \WordToVec{} (of which both variants score only slightly worse than \SERT{}), \LDA{} and \LSI{}. The count-based distributional methods (\LSI{}, \LDA{}) perform better as the dimensionality of the representations increases. This is contrary to \SERT{}, where retrieval performance is very stable across dimensionalities. Interestingly, \DocToVec{} performs very poorly at reconstructing the co-author graph and is even surpassed by the \GraphPCA{} baseline. This is likely due to the fact that \DocToVec{} is trained on expert profiles and is not explicitly presented with document-expert associations. The difference in performance between \DocToVec{} and \SERT{} for \RQRef{2} reflects a difference in architecture: while \SERT{} is directly optimized to discriminate between entities, \DocToVec{} models entities as context in addition to language. Hence, similarities and dissimilarities between entities are preserved much better by \SERT{}.

We answer our second research question as follows. Latent text-based entity representations do encode information about entity relations. However, there is a large difference in the performance of different methods. \SERT{} seems to encode the entity co-associations better than other methods, by achieving the highest performance independent of the vector space dimensionality.
}

\subsection{Analysis of the expert prior in \SERT{}}

\begin{figure*}[!th]
    \centering
    \includegraphics[width=0.95\textwidth]{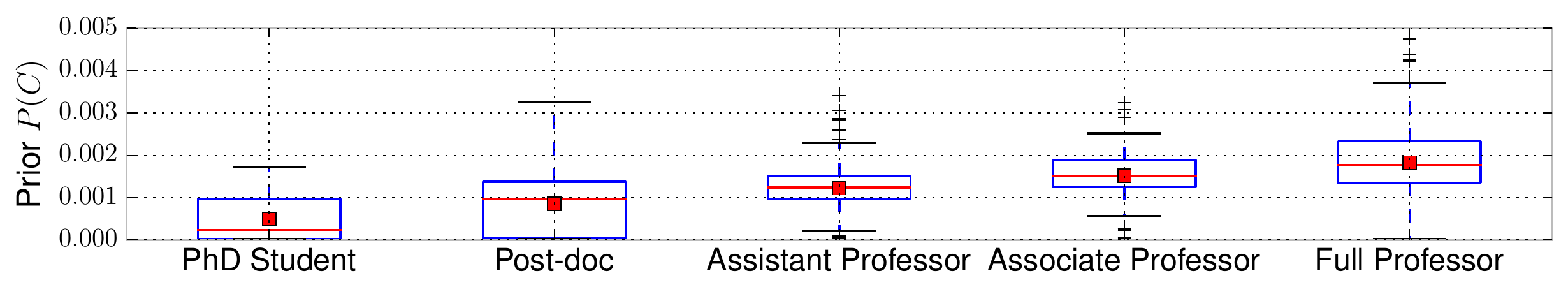}
    \bigskip
    \caption{Box plots of prior probabilities learned by \SERT{}, grouped by the experts' academic rank, for the \BenchmarkTU{} collection. We only show the prior learned for a \SERT{} model with $\VectorDim{} = 32$, as the distributions of models with a different representation dimensionality are qualitatively similar.\label{fig:bias}}
    \smallskip
\end{figure*}

\noindent%
One of the semantic models that we consider, \SERT{}, learns a prior $\Prob{\Candidates{}}$ over entities. The remaining representation learning methods do not encode an explicit entity prior. It might be possible to extract a prior from generic entity vector spaces, e.g., by examining the deviation from the mean representation for every entity. However, developing such prior extraction methods is a topic of study by itself and is out of scope for this paper.

In the case of expert finding, this prior probability encodes a belief over experts without observing any evidence (i.e., query terms in \SERT{}). Which structural information does this prior capture? We now investigate the regularities encoded within this prior and link it back to the hierarchy among scholars in the Tilburg University collection. We estimate a \SERT{} model on the whole \BenchmarkTU{} collection and extract the prior probabilities:
\begin{equation}
\Prob{\Candidates{} = \Candidate{}_i} = \frac{\exp{(\CandidateBias{}_i)}}{\sum_l \exp{(\CandidateBias{}_l)}},
\end{equation}
where $\CandidateBias{}$ is the bias vector of the \SERT{} model in Eq.~\ref{eq:expert-loglinear}.

For \numprint{666} out of \numprint{977} experts in the \BenchmarkTU{} collection we have ground truth information regarding their academic rank \cite{Berendsen2013}.\footnote{126 PhD Students, 49 Postdoctoral Researchers, 210 Assistant Professors, 89 Associate Professors and 190 Full Professors; we filtered out academic ranks that only occur once in the ground-truth, namely Scientific Programmer and Research Coordinator.} Figure~\ref{fig:bias} shows box plots of the prior probabilities, learned automatically by the \SERT{} model from only text and associations, grouped by academic rank. Interestingly, the prior seems to encode the hierarchy amongst scholars at Tilburg University, e.g., Post-docs are ranked higher than PhD students. This is not surprising as it is quite likely that higher-ranked scholars have more associated documents.

The prior over experts in \SERT{} encodes rank within organizations. As mentioned earlier, this is not surprising, as experts (i.e., academics in this experiment) of higher rank tend to occur more frequently in the expert collection. This observation unveils interesting insights about the expert finding task and consequently models targeted at solving it. Unlike unsupervised ad-hoc document retrieval where we assume a uniform prior and normalized document lengths, the prior over experts in the expert finding task is of much greater importance. In addition, we can use this insight to gain a better understanding of the formal language models for expertise retrieval \citep{Balog2006}. \citet{Balog2006} find that, for the expert finding task, the document-oriented language model performs better than a entity-oriented language model. However, the document-oriented model \citep{Balog2006} will rank experts with more associated documents higher than experts with few associated documents. On the contrary, the entity-oriented model of \citet{Balog2006}, imposes a uniform prior over experts. \SERT{} is an entity-oriented model and performs better than the formal document-oriented language model \citep{VanGysel2016experts}. This is likely due to the fact that \SERT{} learns an empirical prior over entities instead of making an assumption of uniformity, in addition to its entity-oriented perspective.

In the case of general entity finding, the importance of the number of associated documents might be of lesser importance. Other sources of prior information, such as link analysis \citep{Page1999}, recency \citep{Graus2016dynamic} and user interactions \citep{Schuth2016forum}, can be a better way of modeling entity importance than the length of entity descriptions.

\vfill\break
%

\section{Conclusions}

In this work we have investigated the structural regularities contained within latent text-based entity representations. Entity representations were constructed from expert finding collections using methods from distributional semantics (\LSI{}), topic models (\LDA{}) and neural networks (\WordToVec{}, \DocToVec{} and \SERT{}). For \LSI{}, \LDA{} and \WordToVec{}, document-level representations were transformed to the entity scope according to the framework of \citet{Demartini2009}. In the case of \DocToVec{} and \SERT{}, entity representations were learned directly. In addition to representations estimated only from text, we considered non-textual baselines, such as:
\begin{inparaenum}[(1)]
	\item random representations sampled from a Normal distribution, and
	\item the rows of the dimensionality-reduced adjacency matrix of the co-association graph.
\end{inparaenum}

We have found that text-based entity representations can be used to discover groups inherent to an organization. We have clustered entity representations using $K$-means and compared the obtained clusters with a ground-truth partitioning. No information about the organization is presented to the algorithms. Instead, these regularities are extracted by the documents associated with entities and published within the organization. Furthermore, we have evaluated the capacity of text-based expert representations to encode co-associations by casting the problem as a ranking task. We discover that text-based representations retain co-associations up to different extents. In particular, we find that \SERT{} entity representations encode the co-association graph better than the other representation learning methods. We conclude that this is due to the fact that \SERT{} representations are directly optimized to discriminate between entities. Lastly, we have shown that the prior probabilities learned by semantic models encode further structural information. That is, we find that the prior probability over experts (i.e., members of an academic institution), learned as part of a \SERT{} model, encodes academic rank. In addition, we discuss the similarities between \SERT{} and the document-oriented language model \citep{Balog2006} and find that the document association prior plays an important role in expert finding.

Our findings have shown insight into how different text-based entity representation methods behave in various applications. In particular, we find that the manner in which entity-document associations are encoded plays an important role. That is, representation learning methods that directly optimize the representation of the entity seem to perform best. When considering different neural representation learning models (\DocToVec{} and \SERT{}), we find that their difference in architecture allows them to encode different regularities: \DocToVec{} models an entity as context in addition to language, whereas \SERT{} learns to discriminate between entities given their language. Thus, \DocToVec{} can more adequately model the topical nature of entities, while \SERT{} more closely captures the similarities and dissimilarities between entities. In the case of expert finding, we find that the amount of textual data associated with an expert is a principal measure of expert importance.

Future work includes the use of text-based entity representations in end-to-end applications. For example, in social networks these methods can be applied to cluster users in addition to network features \citep{VanGysel2015presence,VanGysel2014thesis}, or to induce graphs based on thread participation or hashtag usage. In addition, text-based entity representations can be used as item feature vectors in recommendation systems. Beyond text-only entity collections, there is also a plenitude of applications where entity relations are available. While there has been some work on learning latent representations from entity relations \citep{Bordes2011,Zhao2015}, there has not been much attention given to combining textual evidence and entity relations. Therefore, we identify two additional directions for future work. First, an analysis showing in what capacity entity representations estimated from text alone encode entity-entity relations (beyond the co-associations considered in this work). Secondly, the incorporation of entity-entity similarity in the construction of latent entity representations.

\begin{acks}
We would like to thank the anonymous reviewers for their valuable comments and suggestions.
This research was supported by
Ahold Delhaize,
Amsterdam Data Science,
the Bloomberg Research Grant program,
the Criteo Faculty Research Award program,
the Dutch national program COMMIT,
Elsevier,
the European Community's Seventh Framework Programme (FP7/2007-2013) under
grant agreement nr 312827 (VOX-Pol),
the Google Faculty Research Award scheme,
the Microsoft Research Ph.D.\ program,
the Netherlands Institute for Sound and Vision,
the Netherlands Organisation for Scientific Research (NWO)
under pro\-ject nrs
612.001.116, 
HOR-11-10, 
CI-14-25, 
652.\-002.\-001, 
612.\-001.\-551, 
652.\-001.\-003, 
and
Yandex.
All content represents the opinion of the authors, which is not necessarily shared or endorsed by their respective employers and/or sponsors.
\end{acks}

\vfill\break
\bibliographystyle{ACM-Reference-Format}
\bibliography{ictir2017-expert-regularities}


\begin{thebibliography}{00}


\ifx \showCODEN    \undefined \def \showCODEN     #1{\unskip}     \fi
\ifx \showDOI      \undefined \def \showDOI       #1{{\tt DOI:}\penalty0{#1}\ }
  \fi
\ifx \showISBNx    \undefined \def \showISBNx     #1{\unskip}     \fi
\ifx \showISBNxiii \undefined \def \showISBNxiii  #1{\unskip}     \fi
\ifx \showISSN     \undefined \def \showISSN      #1{\unskip}     \fi
\ifx \showLCCN     \undefined \def \showLCCN      #1{\unskip}     \fi
\ifx \shownote     \undefined \def \shownote      #1{#1}          \fi
\ifx \showarticletitle \undefined \def \showarticletitle #1{#1}   \fi
\ifx \showURL      \undefined \def \showURL       #1{#1}          \fi
\providecommand\bibfield[2]{#2}
\providecommand\bibinfo[2]{#2}
\providecommand\natexlab[1]{#1}
\providecommand\showeprint[2][]{arXiv:#2}

\bibitem[\protect\citeauthoryear{Amig{\'o}, Gonzalo, Artiles, and
  Verdejo}{Amig{\'o} et~al\mbox{.}}{2009}]%
        {Amigo2009}
\bibfield{author}{\bibinfo{person}{Enrique Amig{\'o}}, \bibinfo{person}{Julio
  Gonzalo}, \bibinfo{person}{Javier Artiles}, {and} \bibinfo{person}{Felisa
  Verdejo}.} \bibinfo{year}{2009}\natexlab{}.
\newblock \showarticletitle{A comparison of extrinsic clustering evaluation
  metrics based on formal constraints}.
\newblock \bibinfo{journal}{{\em Information Retrieval\/}}
  \bibinfo{volume}{12}, \bibinfo{number}{4} (\bibinfo{year}{2009}),
  \bibinfo{pages}{461--486}.
\newblock
\showISSN{1386-4564}


\bibitem[\protect\citeauthoryear{Balog, Azzopardi, and de~Rijke}{Balog
  et~al\mbox{.}}{2006}]%
        {Balog2006}
\bibfield{author}{\bibinfo{person}{Krisztian Balog}, \bibinfo{person}{Leif
  Azzopardi}, {and} \bibinfo{person}{Maarten de Rijke}.}
  \bibinfo{year}{2006}\natexlab{}.
\newblock \showarticletitle{{Formal models for expert finding in enterprise
  corpora}}. In \bibinfo{booktitle}{{\em SIGIR}}. \bibinfo{pages}{43--50}.
\newblock


\bibitem[\protect\citeauthoryear{Balog and de~Rijke}{Balog and
  de~Rijke}{2007a}]%
        {Balog2007}
\bibfield{author}{\bibinfo{person}{Krisztian Balog} {and}
  \bibinfo{person}{Maarten de Rijke}.} \bibinfo{year}{2007}\natexlab{a}.
\newblock \showarticletitle{Determining Expert Profiles (With an Application to
  Expert Finding)}. In \bibinfo{booktitle}{{\em IJCAI}}.
  \bibinfo{pages}{2657--2662}.
\newblock


\bibitem[\protect\citeauthoryear{Balog and de~Rijke}{Balog and
  de~Rijke}{2007b}]%
        {Balog2007similarexperts}
\bibfield{author}{\bibinfo{person}{Krisztian Balog} {and}
  \bibinfo{person}{Maarten de Rijke}.} \bibinfo{year}{2007}\natexlab{b}.
\newblock \showarticletitle{Finding similar experts}. In
  \bibinfo{booktitle}{{\em SIGIR}}. ACM, \bibinfo{pages}{821--822}.
\newblock


\bibitem[\protect\citeauthoryear{Balog, Fang, de~Rijke, Serdyukov, and
  Si}{Balog et~al\mbox{.}}{2012}]%
        {Balog2012}
\bibfield{author}{\bibinfo{person}{Krisztian Balog}, \bibinfo{person}{Yi Fang},
  \bibinfo{person}{Maarten de Rijke}, \bibinfo{person}{Pavel Serdyukov}, {and}
  \bibinfo{person}{Luo Si}.} \bibinfo{year}{2012}\natexlab{}.
\newblock \showarticletitle{Expertise Retrieval}.
\newblock \bibinfo{journal}{{\em Found. \& Tr. in Information Retrieval\/}}
  \bibinfo{volume}{6}, \bibinfo{number}{2-3} (\bibinfo{year}{2012}),
  \bibinfo{pages}{127--256}.
\newblock


\bibitem[\protect\citeauthoryear{Balog, Serdyukov, and Vries}{Balog
  et~al\mbox{.}}{2010}]%
        {Balog2010entitytrack}
\bibfield{author}{\bibinfo{person}{Krisztian Balog}, \bibinfo{person}{Pavel
  Serdyukov}, {and} \bibinfo{person}{Arjen P.~de Vries}.}
  \bibinfo{year}{2010}\natexlab{}.
\newblock \showarticletitle{Overview of the {TREC} 2010 entity track}. In
  \bibinfo{booktitle}{{\em TREC}}. \bibinfo{publisher}{NIST}.
\newblock


\bibitem[\protect\citeauthoryear{Baroni, Dinu, and Kruszewski}{Baroni
  et~al\mbox{.}}{2014}]%
        {Baroni2014}
\bibfield{author}{\bibinfo{person}{Marco Baroni}, \bibinfo{person}{Georgiana
  Dinu}, {and} \bibinfo{person}{Germ{\'a}n Kruszewski}.}
  \bibinfo{year}{2014}\natexlab{}.
\newblock \showarticletitle{Don't count, predict! {A} systematic comparison of
  context-counting vs. context-predicting semantic vectors.}. In
  \bibinfo{booktitle}{{\em ACL}}. \bibinfo{pages}{238--247}.
\newblock


\bibitem[\protect\citeauthoryear{Bengio, Ducharme, Vincent, and Janvin}{Bengio
  et~al\mbox{.}}{2003}]%
        {Bengio2003}
\bibfield{author}{\bibinfo{person}{Yoshua Bengio}, \bibinfo{person}{R\'{e}jean
  Ducharme}, \bibinfo{person}{Pascal Vincent}, {and} \bibinfo{person}{Christian
  Janvin}.} \bibinfo{year}{2003}\natexlab{}.
\newblock \showarticletitle{A Neural Probabilistic Language Model}.
\newblock \bibinfo{journal}{{\em JMLR\/}}  \bibinfo{volume}{3}
  (\bibinfo{year}{2003}), \bibinfo{pages}{1137--1155}.
\newblock


\bibitem[\protect\citeauthoryear{Berendsen, de~Rijke, Balog, Bogers, and
  van~den Bosch}{Berendsen et~al\mbox{.}}{2013}]%
        {Berendsen2013}
\bibfield{author}{\bibinfo{person}{Richard Berendsen}, \bibinfo{person}{Maarten
  de Rijke}, \bibinfo{person}{Krisztian Balog}, \bibinfo{person}{Toine Bogers},
  {and} \bibinfo{person}{Antal van~den Bosch}.}
  \bibinfo{year}{2013}\natexlab{}.
\newblock \showarticletitle{On the Assessment of Expertise Profiles}.
\newblock \bibinfo{journal}{{\em JASIST\/}} \bibinfo{volume}{64},
  \bibinfo{number}{10} (\bibinfo{year}{2013}), \bibinfo{pages}{2024--2044}.
\newblock


\bibitem[\protect\citeauthoryear{Blei, Ng, and Jordan}{Blei
  et~al\mbox{.}}{2003}]%
        {Blei2003}
\bibfield{author}{\bibinfo{person}{David~M. Blei}, \bibinfo{person}{Andrew~Y.
  Ng}, {and} \bibinfo{person}{Michael~I. Jordan}.}
  \bibinfo{year}{2003}\natexlab{}.
\newblock \showarticletitle{Latent {Dirichlet} allocation}.
\newblock \bibinfo{journal}{{\em JMLR\/}}  \bibinfo{volume}{3}
  (\bibinfo{year}{2003}), \bibinfo{pages}{993--1022}.
\newblock


\bibitem[\protect\citeauthoryear{Bordes, Weston, Collobert, and Bengio}{Bordes
  et~al\mbox{.}}{2011}]%
        {Bordes2011}
\bibfield{author}{\bibinfo{person}{Antoine Bordes}, \bibinfo{person}{Jason
  Weston}, \bibinfo{person}{Ronan Collobert}, {and} \bibinfo{person}{Yoshua
  Bengio}.} \bibinfo{year}{2011}\natexlab{}.
\newblock \showarticletitle{Learning structured embeddings of knowledge bases}.
  In \bibinfo{booktitle}{{\em AAAI}}.
\newblock


\bibitem[\protect\citeauthoryear{Burges, Shaked, Renshaw, Lazier, Deeds,
  Hamilton, and Hullender}{Burges et~al\mbox{.}}{2005}]%
        {Burges2005}
\bibfield{author}{\bibinfo{person}{Chris Burges}, \bibinfo{person}{Tal Shaked},
  \bibinfo{person}{Erin Renshaw}, \bibinfo{person}{Ari Lazier},
  \bibinfo{person}{Matt Deeds}, \bibinfo{person}{Nicole Hamilton}, {and}
  \bibinfo{person}{Greg Hullender}.} \bibinfo{year}{2005}\natexlab{}.
\newblock \showarticletitle{{Learning to rank using gradient descent}}. In
  \bibinfo{booktitle}{{\em ICML}}. \bibinfo{pages}{89--96}.
\newblock


\bibitem[\protect\citeauthoryear{Caliskan, Bryson, and Narayanan}{Caliskan
  et~al\mbox{.}}{2017}]%
        {Caliskan2017biases}
\bibfield{author}{\bibinfo{person}{Aylin Caliskan}, \bibinfo{person}{Joanna~J.
  Bryson}, {and} \bibinfo{person}{Arvind Narayanan}.}
  \bibinfo{year}{2017}\natexlab{}.
\newblock \showarticletitle{Semantics derived automatically from language
  corpora contain human-like biases}.
\newblock \bibinfo{journal}{{\em Science\/}} \bibinfo{volume}{356},
  \bibinfo{number}{6334} (\bibinfo{year}{2017}), \bibinfo{pages}{183--186}.
\newblock


\bibitem[\protect\citeauthoryear{Clark and Manning}{Clark and Manning}{2016}]%
        {Clark2016improving}
\bibfield{author}{\bibinfo{person}{Kevin Clark} {and}
  \bibinfo{person}{Christopher~D Manning}.} \bibinfo{year}{2016}\natexlab{}.
\newblock \bibinfo{title}{Improving Coreference Resolution by Learning
  Entity-Level Distributed Representations}.
\newblock \bibinfo{howpublished}{arXiv 1606.01323}.   (\bibinfo{year}{2016}).
\newblock


\bibitem[\protect\citeauthoryear{Collobert and Weston}{Collobert and
  Weston}{2008}]%
        {Collobert2008}
\bibfield{author}{\bibinfo{person}{Ronan Collobert} {and}
  \bibinfo{person}{Jason Weston}.} \bibinfo{year}{2008}\natexlab{}.
\newblock \showarticletitle{A unified architecture for natural language
  processing: Deep neural networks with multitask learning}. In
  \bibinfo{booktitle}{{\em ICML}}. \bibinfo{pages}{160--167}.
\newblock


\bibitem[\protect\citeauthoryear{Craswell, de~Vries, and Soboroff}{Craswell
  et~al\mbox{.}}{2005}]%
        {Craswell2005}
\bibfield{author}{\bibinfo{person}{Nick Craswell}, \bibinfo{person}{Arjen~P. de
  Vries}, {and} \bibinfo{person}{Ian Soboroff}.}
  \bibinfo{year}{2005}\natexlab{}.
\newblock \showarticletitle{Overview of the {TREC} 2005 Enterprise Track}. In
  \bibinfo{booktitle}{{\em TREC}}.
\newblock


\bibitem[\protect\citeauthoryear{Davenport and Prusak}{Davenport and
  Prusak}{1998}]%
        {Davenport1998}
\bibfield{author}{\bibinfo{person}{Thomas~H. Davenport} {and}
  \bibinfo{person}{Laurence Prusak}.} \bibinfo{year}{1998}\natexlab{}.
\newblock \bibinfo{booktitle}{{\em Working knowledge: How organizations manage
  what they know}}.
\newblock \bibinfo{publisher}{Harvard Business Press}.
\newblock


\bibitem[\protect\citeauthoryear{Deerwester, Dumais, and Harshman}{Deerwester
  et~al\mbox{.}}{1990}]%
        {Deerwester1990}
\bibfield{author}{\bibinfo{person}{Scott~C. Deerwester},
  \bibinfo{person}{Susan~T. Dumais}, {and} \bibinfo{person}{Richard~A.
  Harshman}.} \bibinfo{year}{1990}\natexlab{}.
\newblock \showarticletitle{Indexing by latent semantic analysis}.
\newblock \bibinfo{journal}{{\em Journal of the American Society for
  Information Science\/}} \bibinfo{volume}{41}, \bibinfo{number}{6}
  (\bibinfo{year}{1990}), \bibinfo{pages}{391--407}.
\newblock


\bibitem[\protect\citeauthoryear{Demartini, Gaugaz, and Nejdl}{Demartini
  et~al\mbox{.}}{2009}]%
        {Demartini2009}
\bibfield{author}{\bibinfo{person}{Gianluca Demartini}, \bibinfo{person}{Julien
  Gaugaz}, {and} \bibinfo{person}{Wolfgang Nejdl}.}
  \bibinfo{year}{2009}\natexlab{}.
\newblock \showarticletitle{A vector space model for ranking entities and its
  application to expert search}.
\newblock In \bibinfo{booktitle}{{\em ECIR}}. \bibinfo{publisher}{Springer},
  \bibinfo{pages}{189--201}.
\newblock


\bibitem[\protect\citeauthoryear{Deng, He, and Gao}{Deng et~al\mbox{.}}{2013}]%
        {Deng2013}
\bibfield{author}{\bibinfo{person}{Li Deng}, \bibinfo{person}{Xiaodong He},
  {and} \bibinfo{person}{Jianfeng Gao}.} \bibinfo{year}{2013}\natexlab{}.
\newblock \showarticletitle{{Deep stacking networks for information
  retrieval}}. In \bibinfo{booktitle}{{\em ICASSP}}.
  \bibinfo{pages}{3153--3157}.
\newblock


\bibitem[\protect\citeauthoryear{Dietz, Kotov, and Meij}{Dietz
  et~al\mbox{.}}{2016}]%
        {Dietz2016tutorial}
\bibfield{author}{\bibinfo{person}{Laura Dietz}, \bibinfo{person}{Alexander
  Kotov}, {and} \bibinfo{person}{Edgar Meij}.} \bibinfo{year}{2016}\natexlab{}.
\newblock \showarticletitle{Utilizing Knowledge Bases in Text-centric
  Information Retrieval}. In \bibinfo{booktitle}{{\em ICTIR}}.
  \bibinfo{publisher}{ACM}, \bibinfo{pages}{5--5}.
\newblock


\bibitem[\protect\citeauthoryear{Fang, Si, and Mathur}{Fang
  et~al\mbox{.}}{2010}]%
        {Fang2010}
\bibfield{author}{\bibinfo{person}{Yi Fang}, \bibinfo{person}{Luo Si}, {and}
  \bibinfo{person}{Aditya~P. Mathur}.} \bibinfo{year}{2010}\natexlab{}.
\newblock \showarticletitle{Discriminative models of integrating document
  evidence and document-candidate associations for expert search}. In
  \bibinfo{booktitle}{{\em SIGIR}}. ACM, \bibinfo{pages}{683--690}.
\newblock


\bibitem[\protect\citeauthoryear{Graus, Tsagkias, Weerkamp, Meij, and
  de~Rijke}{Graus et~al\mbox{.}}{2016}]%
        {Graus2016dynamic}
\bibfield{author}{\bibinfo{person}{David Graus}, \bibinfo{person}{Manos
  Tsagkias}, \bibinfo{person}{Wouter Weerkamp}, \bibinfo{person}{Edgar Meij},
  {and} \bibinfo{person}{Maarten de Rijke}.} \bibinfo{year}{2016}\natexlab{}.
\newblock \showarticletitle{Dynamic collective entity representations for
  entity ranking}. In \bibinfo{booktitle}{{\em WSDM}}. ACM,
  \bibinfo{pages}{595--604}.
\newblock


\bibitem[\protect\citeauthoryear{He, Liu, Li, Zhou, Zhang, and Wang}{He
  et~al\mbox{.}}{2013}]%
        {He2013learning}
\bibfield{author}{\bibinfo{person}{Zhengyan He}, \bibinfo{person}{Shujie Liu},
  \bibinfo{person}{Mu Li}, \bibinfo{person}{Ming Zhou},
  \bibinfo{person}{Longkai Zhang}, {and} \bibinfo{person}{Houfeng Wang}.}
  \bibinfo{year}{2013}\natexlab{}.
\newblock \showarticletitle{Learning Entity Representation for Entity
  Disambiguation.}. In \bibinfo{booktitle}{{\em ACL}}. \bibinfo{pages}{30--34}.
\newblock


\bibitem[\protect\citeauthoryear{Hinton}{Hinton}{1986}]%
        {Hinton1986}
\bibfield{author}{\bibinfo{person}{Geoffrey~E. Hinton}.}
  \bibinfo{year}{1986}\natexlab{}.
\newblock \showarticletitle{Learning distributed representations of concepts}.
  In \bibinfo{booktitle}{{\em 8th Annual Conference of the Cognitive Science
  Society}}, Vol.~\bibinfo{volume}{1}. \bibinfo{address}{Amherst, MA},
  \bibinfo{pages}{12}.
\newblock


\bibitem[\protect\citeauthoryear{Hofmann}{Hofmann}{1999}]%
        {Hofmann1999}
\bibfield{author}{\bibinfo{person}{Thomas Hofmann}.}
  \bibinfo{year}{1999}\natexlab{}.
\newblock \showarticletitle{Probabilistic latent semantic indexing}. In
  \bibinfo{booktitle}{{\em SIGIR}}. ACM, \bibinfo{pages}{50--57}.
\newblock


\bibitem[\protect\citeauthoryear{Huang, Urbana, He, Gao, Deng, Acero, and
  Heck}{Huang et~al\mbox{.}}{2013}]%
        {Huang2013}
\bibfield{author}{\bibinfo{person}{Po-sen Huang},
  \bibinfo{person}{N~Mathews~Ave Urbana}, \bibinfo{person}{Xiaodong He},
  \bibinfo{person}{Jianfeng Gao}, \bibinfo{person}{Li Deng},
  \bibinfo{person}{Alex Acero}, {and} \bibinfo{person}{Larry Heck}.}
  \bibinfo{year}{2013}\natexlab{}.
\newblock \showarticletitle{Learning Deep Structured Semantic Models for Web
  Search using Clickthrough Data}. In \bibinfo{booktitle}{{\em CIKM}}.
  \bibinfo{pages}{2333--2338}.
\newblock


\bibitem[\protect\citeauthoryear{Kenter and de~Rijke}{Kenter and
  de~Rijke}{2015}]%
        {Kenter2015}
\bibfield{author}{\bibinfo{person}{Tom Kenter} {and} \bibinfo{person}{Maarten
  de Rijke}.} \bibinfo{year}{2015}\natexlab{}.
\newblock \showarticletitle{Short text similarity with word embeddings}. In
  \bibinfo{booktitle}{{\em CIKM}}. ACM, \bibinfo{pages}{1411--1420}.
\newblock


\bibitem[\protect\citeauthoryear{Le and Mikolov}{Le and Mikolov}{2014}]%
        {Le2014}
\bibfield{author}{\bibinfo{person}{Quoc~V Le} {and} \bibinfo{person}{Tomas
  Mikolov}.} \bibinfo{year}{2014}\natexlab{}.
\newblock \showarticletitle{Distributed Representations of Sentences and
  Documents}. In \bibinfo{booktitle}{{\em ICML}}. \bibinfo{pages}{1188--1196}.
\newblock


\bibitem[\protect\citeauthoryear{Levy, Goldberg, and Dagan}{Levy
  et~al\mbox{.}}{2015}]%
        {Levy2015}
\bibfield{author}{\bibinfo{person}{Omer Levy}, \bibinfo{person}{Yoav Goldberg},
  {and} \bibinfo{person}{Ido Dagan}.} \bibinfo{year}{2015}\natexlab{}.
\newblock \showarticletitle{Improving distributional similarity with lessons
  learned from word embeddings}.
\newblock \bibinfo{journal}{{\em TACL\/}}  \bibinfo{volume}{3}
  (\bibinfo{year}{2015}), \bibinfo{pages}{211--225}.
\newblock


\bibitem[\protect\citeauthoryear{Levy, Goldberg, and Ramat-Gan}{Levy
  et~al\mbox{.}}{2014}]%
        {Levy2014}
\bibfield{author}{\bibinfo{person}{Omer Levy}, \bibinfo{person}{Yoav Goldberg},
  {and} \bibinfo{person}{Israel Ramat-Gan}.} \bibinfo{year}{2014}\natexlab{}.
\newblock \showarticletitle{Linguistic Regularities in Sparse and Explicit Word
  Representations}. In \bibinfo{booktitle}{{\em CoNLL}}.
  \bibinfo{pages}{171--180}.
\newblock


\bibitem[\protect\citeauthoryear{Li and Xu}{Li and Xu}{2014}]%
        {Li2014}
\bibfield{author}{\bibinfo{person}{Hang Li} {and} \bibinfo{person}{Jun Xu}.}
  \bibinfo{year}{2014}\natexlab{}.
\newblock \showarticletitle{Semantic Matching in Search}.
\newblock \bibinfo{journal}{{\em Found. \& Tr. in Information Retrieval\/}}
  \bibinfo{volume}{7}, \bibinfo{number}{5} (\bibinfo{date}{June}
  \bibinfo{year}{2014}), \bibinfo{pages}{343--469}.
\newblock


\bibitem[\protect\citeauthoryear{Li, Schijvenaars, and de~Rijke}{Li
  et~al\mbox{.}}{2017}]%
        {Li2017academic}
\bibfield{author}{\bibinfo{person}{Xinyi Li}, \bibinfo{person}{Bob
  Schijvenaars}, {and} \bibinfo{person}{Maarten de Rijke}.}
  \bibinfo{year}{2017}\natexlab{}.
\newblock \showarticletitle{Investigating queries and search failures in
  academic search}.
\newblock \bibinfo{journal}{{\em Information Processing \& Management\/}}
  \bibinfo{volume}{53}, \bibinfo{number}{3} (\bibinfo{date}{May}
  \bibinfo{year}{2017}), \bibinfo{pages}{666--683}.
\newblock


\bibitem[\protect\citeauthoryear{Liu}{Liu}{2011}]%
        {Liu2011}
\bibfield{author}{\bibinfo{person}{Tie-Yan Liu}.}
  \bibinfo{year}{2011}\natexlab{}.
\newblock \bibinfo{booktitle}{{\em Learning to Rank for Information
  Retrieval}}.
\newblock \bibinfo{publisher}{Springer}.
\newblock


\bibitem[\protect\citeauthoryear{Luhn}{Luhn}{1958}]%
        {Luhn1958}
\bibfield{author}{\bibinfo{person}{Hans~Peter Luhn}.}
  \bibinfo{year}{1958}\natexlab{}.
\newblock \showarticletitle{The automatic creation of literature abstracts}.
\newblock \bibinfo{journal}{{\em IBM Journal of research and development\/}}
  \bibinfo{volume}{2}, \bibinfo{number}{2} (\bibinfo{year}{1958}),
  \bibinfo{pages}{159--165}.
\newblock


\bibitem[\protect\citeauthoryear{MacQueen}{MacQueen}{1967}]%
        {MacQueen1967}
\bibfield{author}{\bibinfo{person}{James~B. MacQueen}.}
  \bibinfo{year}{1967}\natexlab{}.
\newblock \showarticletitle{Some methods for classification and analysis of
  multivariate observations}. In \bibinfo{booktitle}{{\em Proceedings of the
  Fifth Berkeley Symposium on Mathematical Statistics and Probability, Volume
  1: Statistics}}. \bibinfo{pages}{281--297}.
\newblock


\bibitem[\protect\citeauthoryear{Manning, Raghavan, and Sch\"{u}tze}{Manning
  et~al\mbox{.}}{2008}]%
        {Manning2008ir}
\bibfield{author}{\bibinfo{person}{Christopher~D. Manning},
  \bibinfo{person}{Prabhakar Raghavan}, {and} \bibinfo{person}{Hinrich
  Sch\"{u}tze}.} \bibinfo{year}{2008}\natexlab{}.
\newblock \bibinfo{booktitle}{{\em Introduction to Information Retrieval}}.
\newblock \bibinfo{publisher}{Cambridge University Press},
  \bibinfo{address}{New York, NY, USA}.
\newblock


\bibitem[\protect\citeauthoryear{Mikolov, Chen, Corrado, and Dean}{Mikolov
  et~al\mbox{.}}{2013a}]%
        {Mikolov2013compositionality}
\bibfield{author}{\bibinfo{person}{Tomas Mikolov}, \bibinfo{person}{Kai Chen},
  \bibinfo{person}{Greg Corrado}, {and} \bibinfo{person}{Jeffrey Dean}.}
  \bibinfo{year}{2013}\natexlab{a}.
\newblock \showarticletitle{Distributed Representations of Words and Phrases
  and their Compositionality}. In \bibinfo{booktitle}{{\em NIPS}}.
  \bibinfo{pages}{3111--3119}.
\newblock


\bibitem[\protect\citeauthoryear{Mikolov, Corrado, Chen, and Dean}{Mikolov
  et~al\mbox{.}}{2013b}]%
        {Mikolov2013word2vec}
\bibfield{author}{\bibinfo{person}{Tomas Mikolov}, \bibinfo{person}{Greg
  Corrado}, \bibinfo{person}{Kai Chen}, {and} \bibinfo{person}{Jeffrey Dean}.}
  \bibinfo{year}{2013}\natexlab{b}.
\newblock \bibinfo{title}{Efficient Estimation of Word Representations in
  Vector Space}.
\newblock \bibinfo{howpublished}{arXiv 1301.3781}.   (\bibinfo{year}{2013}).
\newblock


\bibitem[\protect\citeauthoryear{Mikolov, Yih, and Zweig}{Mikolov
  et~al\mbox{.}}{2013c}]%
        {Mikolov2013regularities}
\bibfield{author}{\bibinfo{person}{Tomas Mikolov}, \bibinfo{person}{Wen-tau
  Yih}, {and} \bibinfo{person}{Geoffrey Zweig}.}
  \bibinfo{year}{2013}\natexlab{c}.
\newblock \showarticletitle{Linguistic Regularities in Continuous Space Word
  Representations}. In \bibinfo{booktitle}{{\em HLT-NAACL}}.
  \bibinfo{pages}{746--751}.
\newblock


\bibitem[\protect\citeauthoryear{Page, Brin, Motwani, and Winograd}{Page
  et~al\mbox{.}}{1999}]%
        {Page1999}
\bibfield{author}{\bibinfo{person}{Lawrence Page}, \bibinfo{person}{Sergey
  Brin}, \bibinfo{person}{Rajeev Motwani}, {and} \bibinfo{person}{Terry
  Winograd}.} \bibinfo{year}{1999}\natexlab{}.
\newblock \bibinfo{booktitle}{{\em The PageRank citation ranking: bringing
  order to the web.}}
\newblock \bibinfo{type}{{T}echnical {R}eport}. \bibinfo{institution}{Stanford
  InfoLab}.
\newblock


\bibitem[\protect\citeauthoryear{Pazzani and Billsus}{Pazzani and
  Billsus}{2007}]%
        {Pazzani2007recsys}
\bibfield{author}{\bibinfo{person}{Michael~J Pazzani} {and}
  \bibinfo{person}{Daniel Billsus}.} \bibinfo{year}{2007}\natexlab{}.
\newblock \showarticletitle{Content-based recommendation systems}.
\newblock In \bibinfo{booktitle}{{\em The adaptive web}}.
  \bibinfo{publisher}{Springer}, \bibinfo{pages}{325--341}.
\newblock


\bibitem[\protect\citeauthoryear{Pennington, Socher, and Manning}{Pennington
  et~al\mbox{.}}{2014}]%
        {Pennington2014}
\bibfield{author}{\bibinfo{person}{Jeffrey Pennington},
  \bibinfo{person}{Richard Socher}, {and} \bibinfo{person}{Christopher~D.
  Manning}.} \bibinfo{year}{2014}\natexlab{}.
\newblock \showarticletitle{{GloVe}: Global Vectors for Word Representation}.
  In \bibinfo{booktitle}{{\em EMNLP}}. \bibinfo{pages}{1532--1543}.
\newblock


\bibitem[\protect\citeauthoryear{Pound, Mika, and Zaragoza}{Pound
  et~al\mbox{.}}{2010}]%
        {Pound2010objectretrieval}
\bibfield{author}{\bibinfo{person}{Jeffrey Pound}, \bibinfo{person}{Peter
  Mika}, {and} \bibinfo{person}{Hugo Zaragoza}.}
  \bibinfo{year}{2010}\natexlab{}.
\newblock \showarticletitle{Ad-hoc Object Retrieval in the Web of Data}. In
  \bibinfo{booktitle}{{\em WWW}}. \bibinfo{publisher}{ACM},
  \bibinfo{pages}{771--780}.
\newblock


\bibitem[\protect\citeauthoryear{Robert, Robert, and Honemann}{Robert
  et~al\mbox{.}}{2011}]%
        {Robert2011rules}
\bibfield{author}{\bibinfo{person}{Henry~Martyn Robert},
  \bibinfo{person}{Sarah~Corbin Robert}, {and} \bibinfo{person}{Daniel~H
  Honemann}.} \bibinfo{year}{2011}\natexlab{}.
\newblock \bibinfo{booktitle}{{\em Robert's rules of order newly revised}}.
\newblock \bibinfo{publisher}{Da Capo Press}.
\newblock


\bibitem[\protect\citeauthoryear{Rumelhart, Hinton, and Williams}{Rumelhart
  et~al\mbox{.}}{1985}]%
        {Rumelhart1985}
\bibfield{author}{\bibinfo{person}{David~E. Rumelhart},
  \bibinfo{person}{Geoffrey~E. Hinton}, {and} \bibinfo{person}{Ronald~J.
  Williams}.} \bibinfo{year}{1985}\natexlab{}.
\newblock \bibinfo{booktitle}{{\em Learning internal representations by error
  propagation}}.
\newblock \bibinfo{type}{{T}echnical {R}eport}. \bibinfo{institution}{DTIC
  Document}.
\newblock


\bibitem[\protect\citeauthoryear{Salakhutdinov and Hinton}{Salakhutdinov and
  Hinton}{2009}]%
        {Salakhutdinov2009}
\bibfield{author}{\bibinfo{person}{Ruslan Salakhutdinov} {and}
  \bibinfo{person}{Geoffrey Hinton}.} \bibinfo{year}{2009}\natexlab{}.
\newblock \showarticletitle{Semantic hashing}.
\newblock \bibinfo{journal}{{\em Int. J. Approximate Reasoning\/}}
  \bibinfo{volume}{50}, \bibinfo{number}{7} (\bibinfo{year}{2009}),
  \bibinfo{pages}{969--978}.
\newblock


\bibitem[\protect\citeauthoryear{Schuth}{Schuth}{2016}]%
        {Schuth2016forum}
\bibfield{author}{\bibinfo{person}{Anne Schuth}.}
  \bibinfo{year}{2016}\natexlab{}.
\newblock \showarticletitle{Search Engines That Learn from Their Users}.
\newblock \bibinfo{journal}{{\em SIGIR Forum\/}} \bibinfo{volume}{50},
  \bibinfo{number}{1} (\bibinfo{date}{June} \bibinfo{year}{2016}),
  \bibinfo{pages}{95--96}.
\newblock


\bibitem[\protect\citeauthoryear{Shen, He, Gao, Deng, and Mesnil}{Shen
  et~al\mbox{.}}{2014}]%
        {Shen2014}
\bibfield{author}{\bibinfo{person}{Yelong Shen}, \bibinfo{person}{Xiaodong He},
  \bibinfo{person}{Jianfeng Gao}, \bibinfo{person}{Li Deng}, {and}
  \bibinfo{person}{Gr\'{e}goire Mesnil}.} \bibinfo{year}{2014}\natexlab{}.
\newblock \showarticletitle{A Latent Semantic Model with Convolutional-Pooling
  Structure for Information Retrieval}. In \bibinfo{booktitle}{{\em CIKM}}.
  \bibinfo{pages}{101--110}.
\newblock


\bibitem[\protect\citeauthoryear{TREC}{TREC}{2008}]%
        {TREC2010}
\bibfield{author}{\bibinfo{person}{TREC}.}
  \bibinfo{year}{2005--2008}\natexlab{}.
\newblock \bibinfo{title}{{Enterprise Track}}.
\newblock   (\bibinfo{year}{2005--2008}).
\newblock


\bibitem[\protect\citeauthoryear{Turian, Ratinov, and Bengio}{Turian
  et~al\mbox{.}}{2010}]%
        {Turian2010}
\bibfield{author}{\bibinfo{person}{Joseph Turian}, \bibinfo{person}{Lev
  Ratinov}, {and} \bibinfo{person}{Yoshua Bengio}.}
  \bibinfo{year}{2010}\natexlab{}.
\newblock \showarticletitle{Word representations: a simple and general method
  for semi-supervised learning}. In \bibinfo{booktitle}{{\em ACL}}.
  \bibinfo{pages}{384--394}.
\newblock


\bibitem[\protect\citeauthoryear{van Dijk, Tsagkias, and de~Rijke}{van Dijk
  et~al\mbox{.}}{2015}]%
        {vanDijk2015}
\bibfield{author}{\bibinfo{person}{David van Dijk}, \bibinfo{person}{Manos
  Tsagkias}, {and} \bibinfo{person}{Maarten de Rijke}.}
  \bibinfo{year}{2015}\natexlab{}.
\newblock \showarticletitle{Early Detection of Topical Expertise in Community
  Question Answering}. In \bibinfo{booktitle}{{\em SIGIR}}.
  \bibinfo{publisher}{ACM}, \bibinfo{pages}{995--998}.
\newblock


\bibitem[\protect\citeauthoryear{Van~Gysel}{Van~Gysel}{2014}]%
        {VanGysel2014thesis}
\bibfield{author}{\bibinfo{person}{Christophe Van~Gysel}.}
  \bibinfo{year}{2014}\natexlab{}.
\newblock \bibinfo{title}{Listening to the Flock - Towards opinion mining
  through data-parallel, semi-supervised learning on social graphs.}
\newblock   (\bibinfo{year}{2014}).
\newblock


\bibitem[\protect\citeauthoryear{Van~Gysel, de~Rijke, and Kanoulas}{Van~Gysel
  et~al\mbox{.}}{2016}]%
        {VanGysel2016products}
\bibfield{author}{\bibinfo{person}{Christophe Van~Gysel},
  \bibinfo{person}{Maarten de Rijke}, {and} \bibinfo{person}{Evangelos
  Kanoulas}.} \bibinfo{year}{2016}\natexlab{}.
\newblock \showarticletitle{Learning Latent Vector Spaces for Product Search}.
  In \bibinfo{booktitle}{{\em CIKM}}. ACM, \bibinfo{pages}{165--174}.
\newblock


\bibitem[\protect\citeauthoryear{Van~Gysel, de~Rijke, and Kanoulas}{Van~Gysel
  et~al\mbox{.}}{2017}]%
        {VanGysel2017SERT}
\bibfield{author}{\bibinfo{person}{Christophe Van~Gysel},
  \bibinfo{person}{Maarten de Rijke}, {and} \bibinfo{person}{Evangelos
  Kanoulas}.} \bibinfo{year}{2017}\natexlab{}.
\newblock \showarticletitle{Semantic Entity Retrieval Toolkit}. In
  \bibinfo{booktitle}{{\em Neu-IR SIGIR Workshop}}.
\newblock


\bibitem[\protect\citeauthoryear{Van~Gysel, de~Rijke, and Worring}{Van~Gysel
  et~al\mbox{.}}{2016}]%
        {VanGysel2016experts}
\bibfield{author}{\bibinfo{person}{Christophe Van~Gysel},
  \bibinfo{person}{Maarten de Rijke}, {and} \bibinfo{person}{Marcel Worring}.}
  \bibinfo{year}{2016}\natexlab{}.
\newblock \showarticletitle{Unsupervised, Efficient and Semantic Expertise
  Retrieval}. In \bibinfo{booktitle}{{\em WWW}}. \bibinfo{publisher}{ACM},
  \bibinfo{pages}{1069--1079}.
\newblock


\bibitem[\protect\citeauthoryear{Van~Gysel, Goethals, and de~Rijke}{Van~Gysel
  et~al\mbox{.}}{2015}]%
        {VanGysel2015presence}
\bibfield{author}{\bibinfo{person}{Christophe Van~Gysel}, \bibinfo{person}{Bart
  Goethals}, {and} \bibinfo{person}{Maarten de Rijke}.}
  \bibinfo{year}{2015}\natexlab{}.
\newblock \showarticletitle{Determining the Presence of Political Parties in
  Social Circles}. In \bibinfo{booktitle}{{\em ICWSM}},
  Vol.~\bibinfo{volume}{2015}. \bibinfo{pages}{690--693}.
\newblock


\bibitem[\protect\citeauthoryear{Vinh, Epps, and Bailey}{Vinh
  et~al\mbox{.}}{2010}]%
        {Vinh2010}
\bibfield{author}{\bibinfo{person}{Nguyen~Xuan Vinh}, \bibinfo{person}{Julien
  Epps}, {and} \bibinfo{person}{James Bailey}.}
  \bibinfo{year}{2010}\natexlab{}.
\newblock \showarticletitle{Information theoretic measures for clusterings
  comparison: Variants, properties, normalization and correction for chance}.
\newblock \bibinfo{journal}{{\em JMLR\/}}  \bibinfo{volume}{11}
  (\bibinfo{year}{2010}), \bibinfo{pages}{2837--2854}.
\newblock


\bibitem[\protect\citeauthoryear{Weber, Schek, and Blott}{Weber
  et~al\mbox{.}}{1998}]%
        {Weber1998}
\bibfield{author}{\bibinfo{person}{Roger Weber}, \bibinfo{person}{Hans-J{\"o}rg
  Schek}, {and} \bibinfo{person}{Stephen Blott}.}
  \bibinfo{year}{1998}\natexlab{}.
\newblock \showarticletitle{A quantitative analysis and performance study for
  similarity-search methods in high-dimensional spaces}. In
  \bibinfo{booktitle}{{\em VLDB}}. \bibinfo{pages}{194--205}.
\newblock


\bibitem[\protect\citeauthoryear{Zhao, Zhiyuan, and Sun}{Zhao
  et~al\mbox{.}}{2015}]%
        {Zhao2015}
\bibfield{author}{\bibinfo{person}{Yu Zhao}, \bibinfo{person}{Liu Zhiyuan},
  {and} \bibinfo{person}{Maosong Sun}.} \bibinfo{year}{2015}\natexlab{}.
\newblock \showarticletitle{Representation learning for measuring entity
  relatedness with rich information}. In \bibinfo{booktitle}{{\em IJCAI}}.
  \bibinfo{pages}{1412--1418}.
\newblock


\end{thebibliography}

\end{document}